# Imaging of director fields in liquid crystals using stimulated Raman scattering microscopy

Taewoo Lee,[1] Haridas Mundoor,[1] Derek G. Gann,[1] Timothy J. Callahan,[1] and Ivan I. Smalyukh[1,2,3*]

[1]Department of Physics and Liquid Crystal Materials Research Center, University of Colorado, Boulder, Colorado 80309, USA
[2]Department of Electrical, Computer and Energy Engineering and Materials Science Engineering Program, University of Colorado, Boulder, Colorado 80309, USA
[3]Renewable and Sustainable Energy Institute, National Renewable Energy Laboratory and University of Colorado, Boulder, Colorado 80309, USA
[*]ivan.smalyukh@colorado.edu

**Abstract:** We demonstrate an approach for background-free three-dimensional imaging of director fields in liquid crystals using stimulated Raman scattering microscopy. This imaging technique is implemented using a single femtosecond pulsed laser and a photonic crystal fiber, providing Stokes and pump frequencies needed to access Raman shifts of different chemical bonds of molecules and allowing for chemically selective and broadband imaging of both pristine liquid crystals and composite materials. Using examples of model three-dimensional structures of director fields, we show that the described technique is a powerful tool for mapping of long-range molecular orientation patterns in soft matter via polarized chemical-selective imaging.

## 1. Introduction

Crystals and liquid crystals (LCs) exhibit long-range orientational ordering of building blocks (such as organic molecules) that gives rise to many interesting and practically useful properties [1]. In uniaxial LCs, the local average molecular orientation is described by the director **n** and its spatial variation by the director field **n**(**r**). Probing the slowly varying three-dimensional (3D) patterns of **n**(**r**) on the spatial scales from submicron to tens of micrometers is important from both fundamental and practical standpoints [1,2] since **n**(**r**) determines LC material properties in all applications and its mapping in 3D allows for using LCs as model systems in studying topological defects and interplay of topologies of fields and surfaces [3]. 3D imaging of **n**(**r**) was originally achieved by means of polarizing-mode fluorescence confocal microscopy that required doping of LCs with specially designed fluorescent dyes [4]. The capabilities of imaging **n**(**r**) in 3D were later expanded by utilizing various modes of nonlinear optical microscopy [5-7], most notably the label-free coherent anti-Stokes Raman scattering polarizing microscopy (CARS-PM). However, the main disadvantage of using CARS to image **n**(**r**), similar to the case of other applications of this technique, is that the detected anti-Stokes signal has a strong contribution from the non-resonant background [8-10], which can become an important issue in many LC imaging applications, e.g., when discriminating contributions from different chemical bonds in various biaxial LCs.

In this work, we describe a labeling-free, chemical-selective method for optical imaging of director fields in LCs using stimulated Raman scattering (SRS) microscopy [11,12] based on a single femtosecond oscillator [13]. In SRS, two beams, pump (at frequency $\omega_p$) and Stokes (at frequency $\omega_s < \omega_p$), combine to amplify the Stokes Raman signal when the difference between $\omega_p$ and $\omega_s$ equals the vibrational frequency ($\omega_{vib}$) of a certain chemical bond in the molecules [Fig. 1(a)]. As a result of the amplification of the Raman signal, the pump beam experiences a loss in its intensity, dubbed "stimulated Raman loss" (SRL), shown in Fig. 1(a) while the Stokes beam experiences a gain in its intensity, dubbed "stimulated Raman gain" (SRG). Since the gain/loss mechanism occurs only when the difference between the frequencies of the pump and Stokes beam equals the molecular vibration frequency, the non-resonant background is significantly reduced compared to that of CARS [14]. We show that the SRL signal is strongly dependent on the orientation of the linear polarization direction of the excitation light with respect to average orientation of chemical bonds that follow **n(r)**. We then utilize this orientational sensitivity of SRL to reconstruct **n(r)** by combining the information obtained from images at different linear polarizations of the excitation light. We utilize model director structures of a focal conic domain (FCD) in a smectic LC and particle-induced **n(r)** in a colloids-LC composite to demonstrate the imaging capabilities of this polarizing-mode imaging technique, which we call SRS polarizing microscopy (SRS-PM).

## 2. Experimental

### 2.1 Experimental setup of the SRS-PM

To demonstrate that SRS-PM imaging of **n(r)** is consistent with that of other nonlinear optical imaging techniques, we have integrated this new imaging modality with multimodal nonlinear optical polarizing microscopy described previously [6]. In the integrated setup shown in Fig. 1(b), a tunable (680-1080 nm) Ti:Sapphire oscillator (Chameleon Ultra II, Coherent, 140 fs, 80 MHz repetition rate) is used as the primary excitation source for the coherent Raman imaging. A pulse at 780 nm is split into two beams and one of them then synchronously pumps a highly-nonlinear, polarization-maintaining photonic crystal fiber (PCF, FemtoWhite-800, NKT Photonics) to generate the supercontinuum, which provides the broad Stokes shift (400~3500 cm$^{-1}$) after long pass filters and is stable enough for SRL imaging, as previously demonstrated in Refs. [6,7]. The two beams are recombined at a long pass filter (BLP01-785R-25, Semrock Inc.) spatially and temporally before directing them into an XY-mirror scanning unit (FV300, Olympus) and an inverted microscope (IX81, Olympus). A Faraday isolator is used to protect the Ti:Sapphire laser from the back-reflection of the PCF, and a combination of a half-wave plate and a Glan-laser polarizer allows to control power and polarization of these beams. The Stokes beam is modulated with an electro-optic modulator (EOM, M350-160 KD*P, Conoptics) at 1.7 MHz by using a square-wave function-generator (DS345, Stanford Research Systems). The modulation amplitude and DC offset of RF-driver (model 275, Conoptics) is adjusted to maximize the SRL signal.

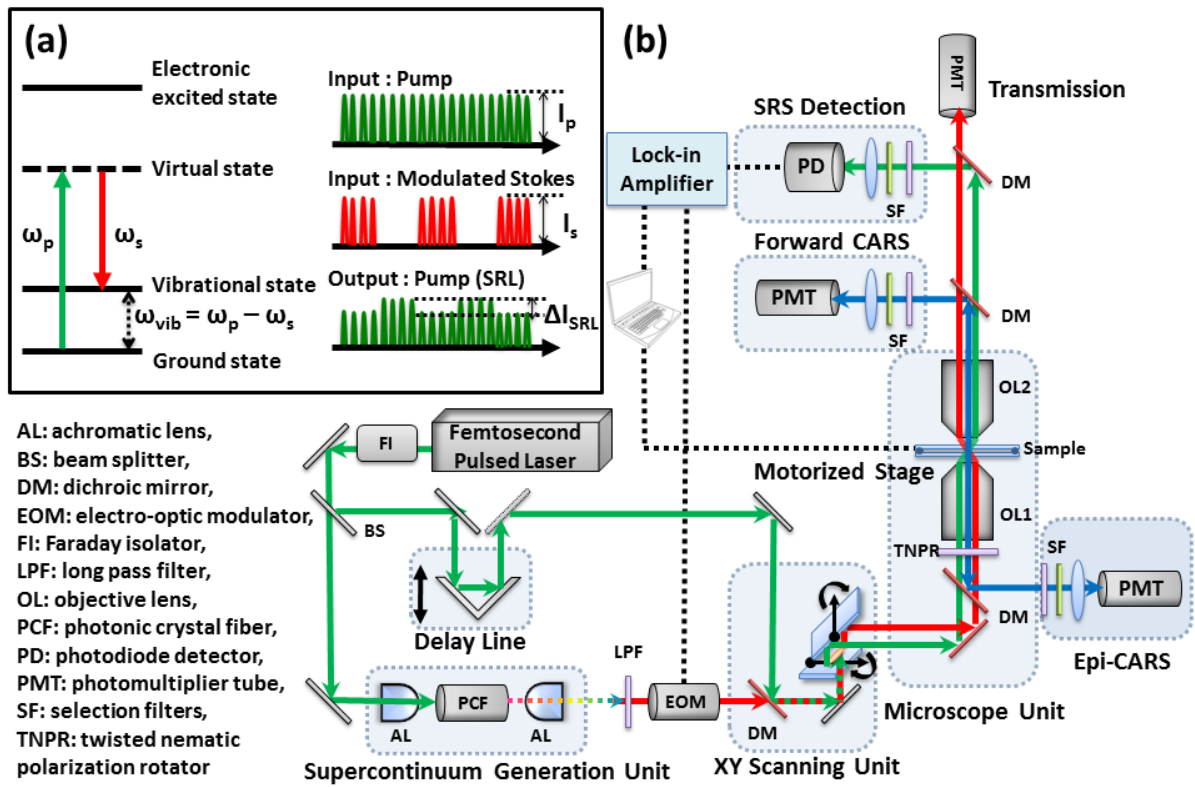

Fig. 1. Principles and a schematic diagram of the SRS-PM setup. (a) Energy diagram of SRS and detection of SRL as the amplitude-modulated component of the pump beam that appears due to the modulated Stokes beam. (b) A schematic diagram of SRS-PM setup integrated with transmission-mode imaging and CARS-PM.

The laser beams are focused into the sample by use of a high numerical aperture (NA) oil-immersion objective lens (UPlanSApo, 100x, NA=1.4, Olympus) and collected by another oil-immersion objective (UPlanApo, 60x, NA=1.42, Olympus) in forward detection mode. The SRL signal is a relatively small part of the pump beam that becomes weakly modulated after passing through the sample. This pump beam is detected by a photodiode and the SRL is selected by means of a lock-in amplifier synchronized with the electro-optic modulator used to modulate the Stokes beam in the excitation part of the setup (Fig. 1). A large-area biased Si photodiode (75.4 mm$^2$, DET100A, Thorlabs) is used for detection of the pump beam at 780 nm. The output from the photodiode passes through a low-pass filter (BLP-1.9+, Mini-Circuit) to suppress the strong laser pulsing signal (80 MHz), before sending it to a high-frequency RF lock-in amplifier (SR844, Stanford Research Systems). The ejected beam from the EOM is detected by the same-type Si photodiode as a reference input for the lock-in amplifier. The analog R (i.e. modulus) or x (i.e. in-phase component)-output of the lock-in amplifier is fed into a modified input of A/D converter, which is synchronized with a motorized microscope sample stage (H117P2IX, Prior). In order to increase signal-to-noise ratio pixel dwell time is set to the range of 100-300 ms and a specified area of the sample is scanned on a pixel by pixel basis with step size 0.2~0.4 μm using a homemade stage control and data acquisition software. In order to reconstruct the SRS-PM images, the SRL signals from output of lock-in amplifier are collected with time correlated pixel information and then converted into a matrix form and are represented to a false color images by using a homemade MatLab program. The CARS-PM signals are detected with photomultiplier tubes (H5784-20, Hamamatsu) in either forward- or epi-detection modes using a series of dichroic mirrors and selection filters.

Raman spectra of the studied LCs are collected by a home-built Raman microscopy setup implemented using an inverted microscope (IX-71, Olympus), SpectraPro-275 spectrometer (Acton Research Corporation), an electron multiplying charge coupled device (EMCCD, iXon3 888, Andor Technology), and a continuous-wave 532-nm laser. The spectra are obtained for linear polarizations of excitation laser light oriented parallel and perpendicular to the far-field director of the planar-aligned LC [Fig. 2(a)]. For the SRS-PM imaging of **n**(**r**), the CN stretching vibration (Raman Shift 2236 cm$^{-1}$) of used LC molecules is chosen for polarization sensitive imaging due to its oscillation parallel to the long molecular axis.

*2.2 Sample preparation*

LC samples were prepared using two glass substrates of thickness 0.17 mm, which were separated by a gap varied within 10–30 μm by incorporating spherical particle spacers of appropriate diameter in-between the substrates. In order to set the surface boundary conditions, inner surfaces of sample cells were coated with polyimide PI-2555 (from HD Microsystems) and then unidirectionally rubbed for planar alignment of **n** along the rubbing direction or treated by a dilute (2 wt%) aqueous solution of N,N-dimethyln-octadecyl-3-aminopropyl-trimethoxysilyl chloride to align **n** perpendicular to the glass plates. To facilitate appearance of FCDs in a SmA LC 8CB (4-cyano-4'-octylbiphenyl, from Frinton Laboratories, Inc.), a hybrid cell is used with one surface treated for planar boundary condition and the other surface treated for perpendicular surface anchoring. The cells were filled with 8CB or nematic LC E7 (from EM Chemicals) by using capillary forces and then sealed by epoxy glue.

### 3. Results

*3.1 Characterization of SRS and conventional Raman signals from aligned LCs*

SRS is a second order nonlinear optical process involving both Pump and Stokes photons. Therefore, the SRS intensity is expected to be proportional to the product of intensities (or powers) of the pump and the Stokes beams, $\Delta I_{SRL} \propto N\sigma_{Raman}P_P P_S$, where $N$ is the number of molecules involved in the SRL process and $\sigma_{Raman}$ is the molecular Raman scattering cross-section. This is consistent with the power dependencies of the SRL signal measured using our SRS-PM experimental setup [Figs. 2(b) and 2(d)]. Orientational ordering of the LC molecules with the CN-bonds along the molecular axes and, on average, along **n**(**r**) yields a strong dependence of the detected SRL signal on the co-linear directions of polarizations of pump and Stokes laser beams with respect to the uniform **n**(**r**) aligned along the rubbing direction [Fig. 2(c)]. This is consistent with ordinary Raman scattering spectra measured for linear polarizations of 532-nm laser excitation light parallel and perpendicular to the director, as shown in Fig. 2(a). The strong orientational sensitivity of SRS-PM signals shown in Fig. 2(c) is instrumental for the capability of 3D imaging of director fields demonstrated below. Although the dependence of $\Delta I_{SRL}$ on $N$ can set detection limits in imaging of many biological systems and composite materials with different types of molecules forming the LC, here we use SRL to probe $\mathbf{n(r)}$ in bulk LC materials composed of 100 % the same molecules. Depolarization of the imaging light due to focusing by high-numerical-aperture objectives and due to the light propagation within the LC currently preclude the possibility of using SRS-PM for the measurement of the scalar order parameter and even mapping of the director fields in thick LC samples [4-7], however, it may be possible to do these types of measurements and imaging in the future provided that the effects of the medium and tight focusing can be compensated using adaptive optical elements such as spatial light modulators [15].

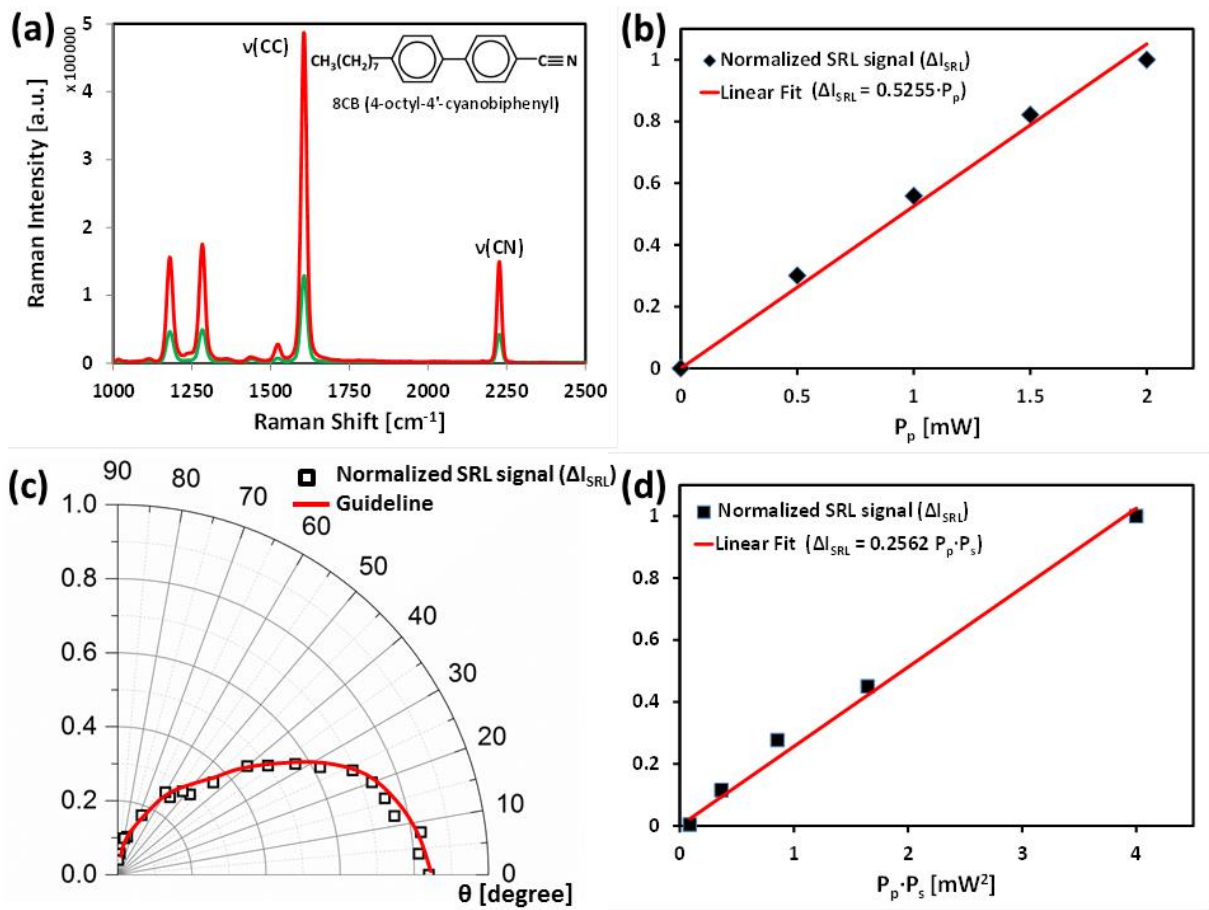

Fig. 2. Characterization of orientation-sensitive Raman and SRL signals. (a) ordinary Raman spectra of an aligned 8CB LC for linear polarizations of excitation light parallel (red) and perpendicular (green) to the far-field director. (b) SRL signal $\Delta I_{SRL}$ vs. pump power ($P_P$) at fixed Stokes power of $P_S$=1.5mW. (c) $\Delta I_{SRL}$ vs. angle $\theta$ between collinear polarizations of excitation beams and $\mathbf{n(r)}$. (d) $\Delta I_{SRL}$ vs. $P_pP_s$. Plots (b-d) were obtained for a planar-aligned E7 LC.

*3.2 SRS-PM imaging of focal conic domains and particle-induced director distortions*

To demonstrate the 3D director imaging capabilities of SRS-PM, we have selected well-defined structures of FCDs [1,4], which represent a class of equidistance-preserving layered configurations. Depth-resolved SRS-PM images obtained for different polarization directions of excitation light are shown in Figs. 3(a)-3(c). Figure 3(d) shows collocated in-plane sections of the FCD obtained using forward-detection CARS-PM imaging for two orthogonal co-linear polarizations of the pump and Stokes beams, similar to that of the SRS-PM image shown in Fig. 3(c). The comparison of the corresponding CARS-PM and SRS-PM images indicates a good agreement of the structures visualized by these two different imaging techniques. This allows one to map out the patterns of molecular orientation and layered structures in 3D based on the SRS-PM images, as demonstrated in Figs. 3(e) and 3(f). In agreement with the reconstructed structures, the equidistant layers of FCDs fold around confocal defect lines, the ellipse and hyperbola and **n**(**r**) within the domain is represented by a series of straight lines connecting these defects, being consistent with the corresponding numerical models.

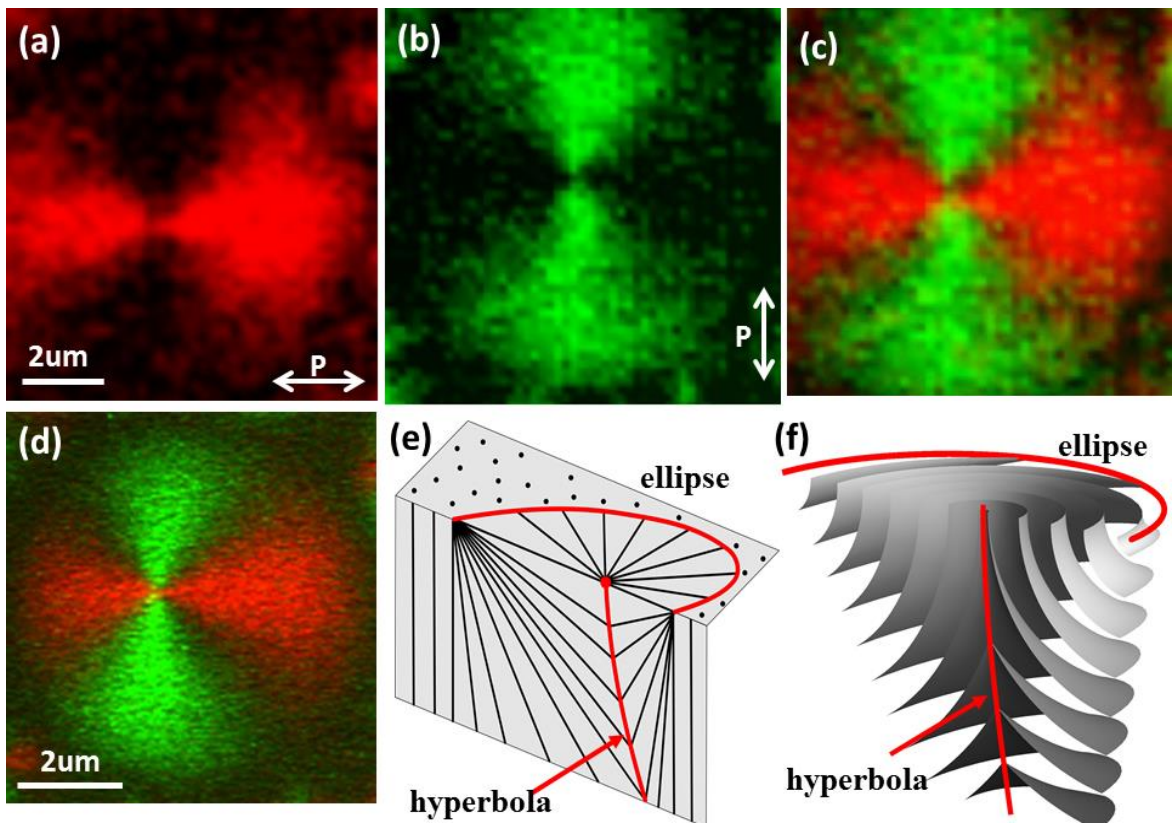

Fig. 3. SRS-PM and CARS-PM images of an FCD in a SmA LC 8CB obtained in the plane of the ellipse. (a-c) Reconstructed SRS-PM images of the FCD for two orthogonal co-linear polarizations of incident pump and Stokes beams (50×50 pixels, 0.2 um pixel size, 100 ms pixel dwell time): (a) horizontal (b) vertical polarization and (c) superimposed images (a) and (b). (d) Collocated CARS-PM images of the same FCD as shown in (c). (e,f) 3D schematics of the FCD containing ellipse and hyperbola defect lines represented by (e) director field **n**(**r**) depicted as black lines, and (f) smectic layers perpendicular to **n**(**r**).

Figure 4 shows an another example of 3D director structure induced by a 3-μm melamine resin colloidal sphere suspended in a planar-aligned 8CB having far-field SmA layers perpendicular to the cell plates and LC molecules far from the particle is aligning along the rubbing direction. The SRS-PM image [Fig. 4(a)] of Smectic LC around a spherical particle, which provides planar surface boundary condition in SmA LC, reveals that **n**(**r**) is along the rubbing direction far from the inclusion but distorted around the spherical particle while satisfying the particle-imposed tangential boundary conditions for the LC molecules. The reconstructed axially symmetric layered pattern contains two singular defect lines emerging from particle's poles along the far-field director, as schematically shown in Fig. 4(b); this layered structure is consistent with our previous studies of particle-induced **n**(**r**) in smectic LCs [6].

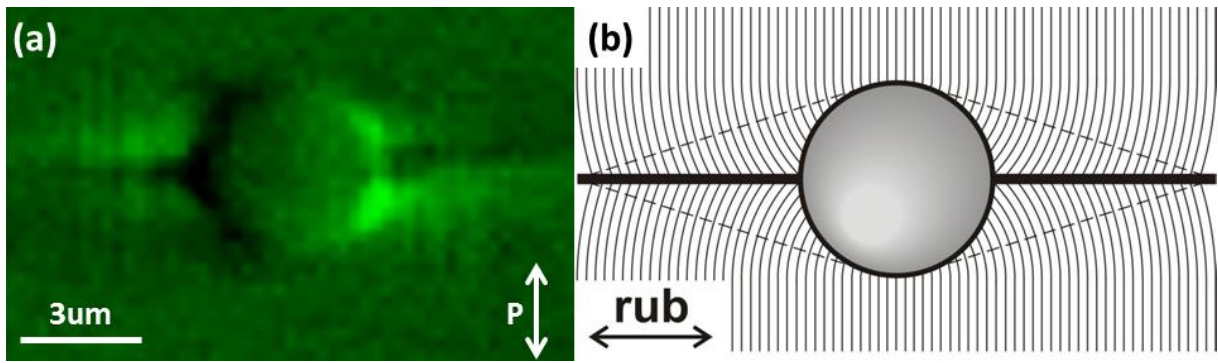

Fig. 4. SRS-PM imaging of an LC-Colloidal composite. (a) SRS-PM image. (b) A schematic of distorted layered structure around a particle in a planar-aligned SmA LC. Thin solid lines show smectic layers; thick straight lines show defects.

No laser-induced director realignment, heating, or phase transitions were observed in the LC sample at laser powers used for this imaging. At pixel dwell times of about 100ms, the excitation laser powers could be as low as 0.5 mW for the pump and 0.5 mW for Stokes at the sample position. However, even when the used laser powers were an order of magnitude higher, the sample scanning during the SRS-PM image acquisition did not alter the director structures in the studied SmA LC, as was confirmed by monitoring the samples in transmission mode using the same setup (Fig. 1). Thus, LC director structures can be imaged by SRS-PM non-destructively and without unintended laser-induced artifacts.

**4. Conclusion**

To conclude, we have demonstrated that SRS-PM is capable of non-invasive 3D imaging of LC director structures. The technique is labeling-free and provides chemically selective orientation-sensitive signals practically free of non-resonant background. SRS-PM may enable direct probing of LC director fields in electro-optic devices, displays, and in various composite systems, such as topological colloids [3]. Selective sensitivity to chemical oscillations will make this type of microscopy especially attractive for imaging of different LC directors in biaxial nematic and smectic LCs by using appropriate chemical bonds of biaxial LC molecules. Furthermore, LC director imaging capabilities may be further expanded by the multiplex coherent Raman microscopy, which was recently implemented by using actively tailored supercontinuum of output from a photonic crystal fiber for a fitted Stokes spectrum via LC-based spatial light modulators [16] to enable simultaneous imaging of different chemical bonds. Future efforts of the SRS-PM development will focus on shortening the pixel dwell time, as needed for imaging of director field dynamics in 3D.

## Acknowledgements

We thank Bohdan Senyuk for his help with sample preparation and Wenlong Yang for discussions. This work was supported by the Institute for Complex Adaptive Matter (H.M.) and NSF grants DMR-0847782 (T.L. and I.I.S.) and DMR-0844115 (H.M. and I.I.S).

## References and links